\documentstyle[12pt]{article}

\newcommand{\eq}{\begin{equation}}
\newcommand{\en}{\end{equation}}
\newcommand{\eqa}{\begin{eqnarray}}
\newcommand{\ena}{\end{eqnarray}}

\newcommand{\NP}[1]{Nucl.\ Phys.\ {\bf #1}}
\newcommand{\PL}[1]{Phys.\ Lett.\ {\bf #1}}
\newcommand{\NC}[1]{Nuovo Cim.\ {\bf #1}}

\newcommand{\MPL}[1]{Mod.\ Phys.\ Lett.\ {\bf #1}}

\begin{document}

\hskip 11.5cm \vbox{\hbox{DFTT 65/92}\hbox{November 1992}}
\vskip 0.4cm
\centerline{\bf KAZAKOV-MIGDAL INDUCED GAUGE THEORY AND THE}
\centerline{\bf COUPLING OF 2D QUANTUM GRAVITY TO D=1 MATTER}
\vskip 1.3cm
\centerline{ M. Caselle, A. D'Adda  and  S. Panzeri}
\vskip .6cm
\centerline{\sl  Dipartimento di Fisica
Teorica dell'Universit\`a di Torino}
\centerline{\sl Istituto Nazionale di Fisica Nucleare,Sezione di Torino}
\centerline{\sl via P.Giuria 1, I-10125 Turin,Italy}
\vskip 2.5cm

\begin{abstract}
Recently Kazakov and Migdal proposed a new approach to the large $N$
limit of SU(N) gauge
theories which could hopefully describe the asymptotically free fixed
point of QCD in 4 dimensions. In this contribution
we review the exact solution of their model
in the case of a $d=1$ compactified lattice and its connection with
the partition function of the vortex-free
sector of the d=1 compactified bosonic string.
Some new results on the addition of a linear term in the K-M model
and on the SU(N) restriction of the general U(N)
model are also presented.
\end{abstract}
\vskip 3.5cm
\hrule
\vskip1.2cm
Talk given By M.Caselle at the LATTICE 92 conference
\vfill
\eject

\newpage

\section{Introduction}

Recently V.Kazakov and A.Migdal proposed a new lattice gauge model (K-M
model in the following), in
which the gauge self-interaction is induced by scalar fields in the
adjoint representation~\cite{KM}.
The action they propose is
defined on a generic  d-dimensional lattice
and has the following form:

\eqa
&S&= \sum_{x} N {\rm Tr} \bigl[m^{2} \phi^{2}(x)\nonumber \\
&-& \sum_{\mu} \phi(x)U_{x,x+\mu}\phi(x+\mu)
U^{\dagger}_{x,x+\mu}\bigr]
\label{km}
\ena
where $\phi(x)$ are Hermitian $N \times N$ matrices defined on the sites
$x$ of the lattice,
and
 $U_{x,x+\mu}$ are Unitary $N \times N$ matrices, defined  on the links
$(x,x+\mu)$, and play the role, as in the usual lattice
discretization of Yang-Mills theories, of the gauge field (see also
ref.~\cite{dalley,h} for related  models).

The main feature of this action is that the  gauge
self-interaction term is  absent thus allowing an exact
solution of the model in the large $N$ limit for
$d>1$~\cite{KM,M,G} and for any value of $N$ in $d=1$~\cite{CAP}.
Moreover a (rather peculiar) gauge self-interaction terms is induced,
 by integration over the matter fields.  The hope is then to have a
theory which is exactly
solvable in the large N limit, and
 in the same universality class of ordinary QCD.
The induced  effective action for the gauge
field, obtained by integrating over the
scalar field $\phi$ is\footnote{Notice in the following equation
 the factor $2$ in front of $m^2$. Such factor has been omitted ,
 presumably due to a misprint, in ref.~\cite{KM} and such mistake has
 propagated to all the following literature that we know of.} :
\eq
S_{ind}[U] = - \frac{1}{2} \sum_{\Gamma} \frac{ |{\rm Tr} U[\Gamma]|^{2}}
{l[\Gamma] (2m^2)^{l[\Gamma]}}~~~,
\label{gauge}
\en
where $l[\Gamma]$ is the length of the loop $\Gamma$, $U[\Gamma]$ is
the ordered product of link matrices along $\Gamma$ and the summation
is over all closed loops.

An independent, interesting feature of this model is its deep
connection with the
theory of non-critical strings.
This connection has been clearly  discussed by
Gross~\cite{G} in the
case of the $d=1$ open chain and
can be made explicit if one integrates over the gauge
fields. The
integration can be performed by using the well known
Harish-Chandra,
 Itzykson and Zuber, Mehta  formula~\cite{hciz}:

\eqa
I(\phi(x),\phi(y))& =& \int D U \exp \left(
N \, tr \phi(x) U \phi(y) U^{\dagger}
\right) \nonumber \\
&\propto& \frac{\det_{ij} \exp(N \lambda_i(x)
\lambda_j(y) )}{\Delta(\lambda(x))
\Delta(\lambda(y))}
\label{izhc}
\ena
where
$\lambda_i(x)$ are the eigenvalues of the matrix $\phi(x)$,
$\Delta(\lambda)$
is the Vandermonde determinant,
\eq
\Delta(\lambda) = \prod_{i<j} (\lambda_i-\lambda_j)
\label{vandermond}
\en
 and $(x,y)$ are nearest neighbour links
of the lattice.

As the gauge variables $U_{x,x+\mu}$ are independent degrees of freedom,
the angular part of the matter fields $\phi(x)$ can be absorbed in them,
and the integral can be performed by means of (\ref{izhc}) leading to a
description purely in terms of the eigenvalues of $\phi(x)$. The same is
not possible in matrix models whenever the matter fields are defined on
lattices containing closed loops.
 It is exactly this kind of obstruction
which doesn't allow,
a description of $d>1$ bosonic strings in terms of the eigenvalues
 only,  and which manifests itself in the case of the
$d=1$ compactified bosonic string as a  vortex contribution~\cite{KG}.
As a consequence of this we expect that
the Kazakov-Migdal model, in the case of a $d=1$ compactified lattice,
should
correspond to the singlet, vortex free, solution of the compactified
bosonic string. We will see below that this is indeed the case and that
the correspondence between the two is rather non trivial.

Despite all these nice features it is now clear that the K-M model
in its original formulation eq.(\ref{km}) cannot induce ordinary QCD.
Let us mention in this respect that
the induced gauge
theory eq.(\ref{gauge})  has a super-confining behaviour~\cite{KSW,KMSW}
since the matter fields are in the
adjoint representation  and
that the large $N$ solution in $d>1$, $m^2>d$ has no continuum
limit~\cite{G}.
Various modifications  of the original K-M action have been
recently proposed~\cite{KMSW,Mig92d,Kh-Ma,ru} to eliminate these
problems, keeping exact solvability in the large $N$ limit.

It would be desirable to have some simple model to enable us to
choose among these proposal
and to discuss their properties, as well as indicating
how good is the $N=\infty$ approximation for finite values of $N$.
Finally one would like to
understand the nature of the phase transition which occurs at the
critical value $m^2_c=d$ of the mass parameter and, above all, what
happens in the weak coupling phase ($m^2<m^2_c$) where, at least in the
original KM model the large $N$ solution
with the ordinary translationally invariant saddle point has been proved
to be unstable~\cite{G}.

An attempt to answer these questions can be made
using the fact that as mentioned above
the K-M model can be solved exactly
for any value of $N$ in
 the case of a $d=1$ compactified lattice made
of $S$ matter fields, for any value of $S$. This
 solution can be  obtained by
using simple combinatoric properties of the permutation group, the key
trick being the reduction of the permutation group to its cyclic
representations as described in~\cite{CAP}. Moreover it is also
 possible  to implement the $det~ U =1$ constraint so as to move
from $U(N)$ to $SU(N)$ theories  while keeping exact
solvability.

It is interesting to see that,
as one would expect from the above discussion, this exact
solution also describes
the vortex-free sector of the $d=1$ compactified bosonic string.
In particular  one can see that in this case the transition at
$m^2=m^2_c=1$ separates the upside-down oscillator phase
 from the standard matrix oscillator description of the $d=1$ bosonic
string, and that the analytic
continuation  in the mass parameter from the strong to the weak coupling
phase provides the correct prescription to obtain the physical
properties of the upside-down oscillators from the standard matrix
oscillators ~\cite{bk}.

This report is organized as follows: in the next section we will discuss
the exact solution in the case of the original K-M model eq.(\ref{km})
and its connection with string theories. Sect.3 will be devoted to the
SU(N) restriction of the model while in the last
section we discuss the $d=1$ induced gauge theory.

\section{ Exact solution of the K-M model on a d=1 compactified lattice}

\noindent
The partition function
of the Kazakov-Migdal model for a $U(N)$ gauge group on a 1d
lattice  with $S$ sites  compactified on a circle, in the region $m^2>1$
is:

\eq
 Z_{U(N)}(q)~=~C_{N,S}\frac{q^{N^2/2}}{(1-q)(1-q^2)\cdots (1-q^N)}
\label{d1}
\en
with $ q= a^{-2S} $ , $a^2=m^2+\sqrt{m^4-1}$ and

\eq
C_{N,S}~ =~\frac{1}{2\pi} \frac{(N!)^{S}}{N^{SN(N-1)/2}} \left(
\frac{\pi}{N} \right)^{\frac{NS}{2}}
\en

The same partition function  was obtained in a completely different
fashion by Boulatov and Kazakov in ref.\cite{bk} as the one describing
 the singlet (vortex free) part of the partition function for a 1d
 string. The only difference with our result
 is that the theory considered in ~\cite{bk} has a
 continuous
compactified target space. In the continuum limit (which implies
$m^2\to m_c^2=1$)  we can
identify, as expected,
 the argument $ q = a^{-2S} $ of our solution
 (\ref{d1})   with $ q = e^{-\beta\omega} $ of~\cite{bk}
 where  $\beta $ is the length of the string and $ \omega$
 is the frequency of the
oscillators.
It is thus clear that we can have  two
completely different phases according to whether the point
 $ m^{2} =1 $ is
reached from above or from below, the former corresponding to a real and
the latter to an imaginary frequency $ \omega $ .
While the $m^2>m^2_c$ phase simply describes a set of $N$ fermions in
an ordinary harmonic potential, the $m^2<m^2_c$ phase, describing an
 upside down oscillator is much more interesting: this is the ``string''
phase of the K-M model.

Notice at this point that all the
 calculation leading to eq.(\ref{d1}) were done in the
regime $ m^2 > 1 $. In such regime the quadratic potential is stable,
$ a $ is real and all integrals are well defined. In the
weak coupling regime ($ m^2 < 1 $) the quadratic potential is unstable
and such instability manifests itself in the divergence of the integrals
over the eigenvalues. This is simply the  $d=1$ case of the
unstability found in the large $N$ solution by Gross in the same regime,
but in this case this is not the end of the story, since (knowing the
exact form of the partition function) we
 can define the partition function for $ m^2 < 1 $
as the analytic continuation from $ m^2 > 1 $.
In terms of the variable $ q = a^{-2S} $ it means an analytic
continuation from the real axis with $ q < 1 $ to the unit circle.
This is absolutely non trivial in the continuum formulation (see for
instance  \cite{bk} where it was achieved by means of the
 the introduction of an SU(N)
invariant cutoff at large $ \lambda$ 's), but it is natural in the
discrete (K-M) theory
where it is possible to analytically
continue from the strong to the weak coupling regime {\sl before}
performing  the continuum limit. In this way one recovers the correct
prescription for the upside-down oscillators, namely that their
properties  are obtained via the substitution $ \omega \to i\omega $ .

It would be interesting to reach this same phase also in $d>1$ but, as
we have seen this requires an exact solution for finite $N$ and finite
lattices. Indeed it is easy to see already in $d=1$   that the saddle
point solution (which in this case selects only the numerator of
(\ref{d1}): $q^{N^2/2}$) cannot see all the singularities
at the denominator
which dominate the partition function
near the critical point.

\section{From U(N) to SU(N) models in d=1}

In the following we shall systematically skip calculations, which are
straightforward extensions of those described in~\cite{CAP} and shall
only comment the results.

The simplest modification of the K-M proposal is the addition of a
linear term: $\sigma\phi(x)$ in the action. If the coefficient $\sigma$
is constant in space the correction to the  K-M partition
function can be evaluated exactly in any
dimension $d$ (simply by shifting the $\phi$ field in the K-M action):
\eq
Z_{U(N)}(\sigma,m,d)=e^{\frac{N^2}{4}\sigma^2\frac{Vol}{m^2-d}}
Z^{(N)}_{K-M}(m,d)
\label{l3}
\en
where $Vol$ is the volume of the lattice.
It is interesting to notice that the addition of a linear term induces
an essential singularity in the partition function as $m^2$ goes to 1.
This term survives the infinite volume and large N limits, and tells us
that the large $N$ solution~\cite{G} is unstable in the case of the
$U(N)$
group also for $m^2>d$ (while in the case of $SU(N)$ the linear term is
explicitly absent due to the traceless condition).

If $d=1$
the partition function can be evaluated
exactly even in the most general case in which the coefficient $\sigma$
of the
linear term is a function of the  position. The action is:

\eqa
&S&= \sum_{x} N {\rm Tr} \bigl[m^{2} \phi^{2}(x)
+\sigma(x)\phi(x)
\nonumber \\
&-& \phi(x)U_{x,x+1}\phi(x+1)
U^{\dagger}_{x,x+1}\bigr]
\label{kmd1}
\ena

and the resulting partition function is:
\eq
Z_{U(N)}(\sigma,q)=e^{\frac{N^2}{2}\sum_x\mu^2(x)}Z_{U(N)}(q)
\label{l1}
\en

where the $\mu$'s are solutions of the chain of equations:
\eq
\sigma(x)=a\mu(x)-\frac{1}{a}\mu(x-1)
\en
and $q$,$a$ and $Z_{U(N)}(q)$ are the same of eq.(\ref{d1}).

This result allows us to implement explicitly the traceless constraint
in the partition function by taking a purely imaginary linear correction
$\sigma(x)\equiv i\gamma(x)$ ($\gamma\in{\bf R}$),
and integrating in each site over $\gamma(x)$.
The resulting partition function is, apart from irrelevant numerical
factors:
\eqa
&Z_{SU(N)}&=(\sqrt{q}-\frac{1}{\sqrt{q}})Z_{U(N)}\nonumber\\
&=&C_{N,S}\frac{q^{(N^2-1)/2}}{(1-q^2)(1-q^3)\cdots (1-q^N)}
\ena

As it might be expected the restriction to $SU(N)$ eliminates from the
partition function the contribution of one of the $N$ free fermions.

\section{Induced gauge action in d=1}
It is also possible in $d=1$ to integrate explicitly
over the matrix fields $\phi $ and obtain the partition
function as a function of the gauge fields only (see~\cite{CAP} for the
details). The result is  :
\eqa
Z(q)&=&
 \int_0^{2\pi}\prod_{k=1}^N\frac{d\theta_k}{2\pi}
\vert \Delta(e^{i\theta})\vert^2\nonumber\\
&&\prod_{k,m=1}^{N}\left[
\frac{q^{1/2}}{1-qe^{i(\theta_{k}-\theta_{m})}}
\right]
\label{c1}
\ena
where, the $\theta_i$'s are the invariant
angles of the product of the unitary matrices along the lattice.

It is interesting to observe that the dependence on the   $\theta_i$'s
in (\ref{c1}) is not affected by the restriction to
 $SU(N)$. This can be understood by noticing that, since the addition
 of a linear term in (\ref{kmd1}) corresponds to a shift of the fields
$\phi(x)$, it does not affect the gaussian integration over the
$\phi$'s.

Since the induced gauge action is known one can address the question of
its equivalence with some known 2d gauge theory (notice that our
lattice is equivalent to a plaquette and any 2d LGT can be reduced to a
1-plaquette problem). In particular one can look to the Villain models
which have a simple expression in terms of invariant angles and
 are also exactly solved for any value of $N$~\cite{onofri}. One can
easily see from eq.(\ref{c1}) that the distribution of the
 eigenvalues $\theta_i$'s is not peaked around $\theta_i = 0$ and it
becomes uniform at the critical point $q \to 1$. That clearly denotes
that the action (\ref{c1}) is not related to the Villain action
 (in fact the action (\ref{c1}) is related, through suitable
identifications of the couplings, to a $ratios$
of Villain partition functions rather than to a Villain model itself).
 This  feature is most probably related to the fact that
 contours of arbitrary length contribute to the induced
gauge action (\ref{gauge})~\cite{KMSW}, a circumstance that has no
 analogue in ordinary gauge theories. This is a further indication that
 suitable modifications of the original K-M proposal are needed.

\end{document}